\documentclass[11pt]{article}
\usepackage{amsmath,amssymb,color,cite}

\newcommand{\rep}[1]{\ensuremath{\mathbf{#1}}}

\usepackage{amsfonts}
\def\oneone{\rlap 1\mkern4mu{\rm l}}

\newcommand{\eq}[1]{(\ref{#1})}

\newcommand{\be}{\begin{equation}}
\newcommand{\ee}{\end{equation}}
\newcommand{\bea}{\setlength\arraycolsep{2pt} \begin{eqnarray}}
\newcommand{\eea}{\end{eqnarray}}
\newcommand{\nn}{\nonumber}

 \csname
@addtoreset\endcsname{equation}{section}

\def\ft#1#2{{\textstyle{\frac{\scriptstyle #1}{\scriptstyle #2} } }}
\def\fft#1#2{{\frac{#1}{#2}}}

\def\0{{\sst{(0)}}}
\def\1{{\sst{(1)}}}
\def\2{{\sst{(2)}}}
\def\3{{\sst{(3)}}}
\def\4{{\sst{(4)}}}
\def\5{{\sst{(5)}}}
\def\6{{\sst{(6)}}}
\def\7{{\sst{(7)}}}
\def\8{{\sst{(8)}}}
\def\sst#1{{\scriptscriptstyle #1}}
\def\oneone{\rlap 1\mkern4mu{\rm l}}

\def\del{{\partial}}

\begin{document}

\thispagestyle{empty}

\begin{flushright}\small
USTC-ICTS-15-08\ \ \ \  MI-TH-1529    \\
%PRE-PR-INT
\end{flushright}

%%%%%%%%%%%%%%%%%%%%%%%%%%%%%%%%%%%%%%%%%

\bigskip
\bigskip

\vskip 10mm

\begin{center}

  {\Large{\bf Membrane Duality Revisited}}

\end{center}

%%%%%%%%%%%%%%%%%%%%%%%%%%%%%%%%%%%%%%%%

\vskip 6mm

\begin{center}

{\bf M.J. Duff$^\ast$, J.X. Lu$^\dag$, R. Percacci$^\ddag$, 
C.N. Pope$^{\S, \diamondsuit}$, H. Samtleben$^\P$ , E. Sezgin$^\S$}

\vskip 4mm

$^\ast$\,{\em Theoretical Physics, Blackett Laboratory\\
Imperial College London, London SW7 2AZ, United Kingdom} \\
\vskip 4mm

$^\dag$\,{\em Interdisciplinary Center for Theoretical Study\\
University of Science and Technology of China, Hefei, Anhui 230026, China} \\
\vskip 4mm

$^\ddag$\,{\em SISSA, via Bonomea 265, Trieste, Italy and INFN, Sezione di Trieste} \\
\vskip 4mm

$^\S$\,{\em George P. and Cynthia W. Mitchell Institute \\for Fundamental
Physics and Astronomy \\
Texas A\&M University, College Station, TX 77843-4242, USA}\\
\vskip 4mm

$^\diamondsuit$\,{\em DAMTP, Centre for Mathematical Sciences,\\
 Cambridge University, Wilberforce Road, Cambridge CB3 OWA, UK}\\
\vskip 4mm
$^\P$\,{\em Universit\'e de Lyon, Laboratoire de Physique, UMR 5672, CNRS et ENS de Lyon,\\
46 all\'ee d'Italie, F-69364 Lyon CEDEX 07, France} \\
\vskip 4mm

\end{center}

\vskip0.5cm

\begin{center} {\bf Abstract } \end{center}

\begin{quotation}\noindent

Just as string T-duality originates from transforming field equations 
into Bianchi identities on the
string worldsheet, so it has been suggested that M-theory 
U-dualities originate from transforming
field equations into Bianchi identities on the membrane worldvolume. 
However, this encounters a problem  unless the target space has dimension $D = p + 1$. We identify the problem to be the nonintegrability of the U-duality 
transformation assigned to the pull-back map.
Just as a double geometry  renders manifest the $O(D,D)$ string T-duality, here we show in 
the case of the M2-brane in $D = 3$ that a generalised geometry renders manifest 
the $SL(3) \times SL(2)$ U-duality.
In the case of M2-brane in $D=4$, with and without extra target space 
coordinates, we show that only the ${\rm GL}(4,R)\ltimes R^4$ subgroup of the expected $SL(5,R)$ 
U-duality symmetry is realised.

\end{quotation}

\newpage
\setcounter{page}{1}

%\tableofcontents

%%%%%%%%%%%%%%%%%%%
\section{Introduction}
%%%%%%%%%%%%%%%%%%%

%%%%%%%%%%%%%%%%%%%
\subsection{The Story So Far}

\subsubsection*{Strings, T-duality and Double Geometry}
\indent
Some time ago  \cite {Duff:1989tf}, it was pointed out that strings moving in an $D$-dimensional 
space $M^D$ with coordinates $X^{\mu}(\tau, \sigma)$, background metric $g_{\mu\nu}(X)$ and 
2-form $B_{\mu\nu}(X)$, could usefully be described by a doubled geometry with $2D$-dimensional 
coordinates
\be
Z^M=(X^{\mu},Y_{\sigma})
\ee
and doubled metric\footnote{$G_{MN}$ had previously appeared in \cite{Giveon:1988tt} with a different 
physical interpretation as a metric on phase space.} 
 \be
G_{MN}= \left(
\begin{array}{cc}
g_{\mu\nu}-B_{\mu\rho}\,g^{\rho\sigma}B_{\sigma\nu}& B_{\mu\rho}\,g^{\rho\sigma}\\
-g^{\mu\sigma}B_{\sigma\nu}&g^{\mu\nu}
\end{array}\,.
\right)
 \ee  
The motivation was twofold; worldsheet and spacetime:
\begin{enumerate}
\item{\bf Worldsheet}

 In the case when $M^D$ is the $D$-torus $T^D$, this renders manifest 
the $O(D,D)$ T-duality by combining worldsheet field equations and 
Bianchi identities via the constraint
\be
\Omega_{MN}\epsilon^{ij}\partial_{j}Z^N=
G_{MN}\sqrt{-\gamma}\gamma^{ij}\partial_{j}Z^N\,,
\label{dcon}
\ee
where
\be
\Omega_{MN}=\left(
\begin{array}{cc}
 0 & \delta_{\mu}{}^{\beta} \\
\delta^{\alpha}{}_{\nu} &0
\end{array}
\right)\,,
\ee
and $\gamma_{ij}$ is the worldsheet metric.  

\item{\bf Spacetime}

 In the case when $M^D$ is a generic manifold,  
the $2D$-dimensional diffeomorphisms with parameter 
$\xi^M=(\xi^{\mu}, \lambda_{\alpha})$
suggest a way of unifying\footnote{An earlier alternative suggestion \cite{Duff:1986ne} was to use 
the non-symmetric metric $g_{\mu\nu}+b_{\mu\nu}$. The two alternatives are related by the two-vielbein 
approach \cite {Siegel:1993xq}.} $D$-dimensional diffeomorphisms 
\be
\delta g_{\mu\nu}=-\partial_{\mu}\xi^{\rho} g_{\rho\nu}-\partial_{\nu} 
\xi^{\rho} g_{\mu\rho}-\partial_{\rho}g_{\mu\nu}\xi^{\rho} \,,
\ee
and 2-form gauge invariance
\be
\delta B_{\mu\nu}=\partial_{\mu}\lambda_{\nu}-\partial_{\nu}\lambda_{\mu}\,.
\ee
After all, $G_{MN}$ is just the Kaluza-Klein metric with 
spacetime metric $g_{\mu\nu}$,
gauge field $A_{\mu}{}^a$ and internal metric $g_{ab}$
 \be
G_{MN}= \left(
\begin{array}{cc}
g_{\mu\nu}+A_{\mu}{}^{a}g_{ab}A_{\nu}{}^b& A_{\mu}{}^{a}g_{ab}\\
g_{ab}A_{\nu}{}^b&g_{ab}
\end{array}
\right)\,,
 \ee  
where the ``gauge field'' is $B_{\mu\alpha}$ and the ``internal'' 
metric is $g^{\alpha\beta}$.  If this programme were successful one 
would expect the $SL(D)/SO(D)$ coset of general relativity to be promoted 
to an $O(D,D) /(SO(D) \times SO(D))$, as conjectured in 
\cite{Duff:1985bv,Duff:1986ne}.
	 
	In summary, the worldsheet goal of rendering manifest the string 
T-duality $O(D,D)$ by doubling the coordinates was achieved successfully 
in \cite {Duff:1989tf} and a T-dual worldsheet action using the 
doubled coordinates was then constructed in \cite{Tseytlin:1990va}. However,
there were missing ingredients in the spacetime approach:  
the generalized diffeomorphisms were subsequently supplied in 
\cite{Siegel:1993xq,Siegel:1993th}
\be
\delta G_{MN}=\xi^P\partial_PG_{MN}+(\partial_M\xi^P-
\partial^P\xi_M)G_{PN}+(\partial_N\xi^P-\partial^P\xi_N)G_{MP}\,,
\ee
and the section condition subsequently supplied in \cite{Hull:2009mi}
\be
\Omega^{MN}\partial_M\partial_N=0\,.
\ee
(The need for the section condition has, however, 
been called into question \cite{Grana:2012rr,Berman:2013cli}.)  
Once these ingredients were included, it was possible also to 
build a generalised spacetime action for $G_{MN}$. This activity came to 
be known as ``Double Field Theory.'' 
\end{enumerate} 

For further developments and variations on this doubled geometry theme, 
in addition to those already cited, 
including ``Generalised geometry'' and the $E_{11}$ approach see, 
for example, \cite{Maharana:1992my,Schwarz:1993vs,Giveon:1994fu,West:2001as,Gualtieri:2003dx,
Hull:2006va,Pacheco:2008ps,Hull:2009mi,Hull:2009zb,Hohm:2010jy,Hohm:2010pp,Hitchin:2010qz,
Berman:2010is,Berman:2011jh,Coimbra:2011ky,Thompson:2011uw,Coimbra:2012af,Hohm:2012gk,
West:2012qz,Grana:2012rr, Hatsuda:2012vm,Hatsuda:2013dya,Aldazabal:2013sca,Andriot:2012vb,
Berman:2013uda,Godazgar:2013bja,Nibbelink:2013zda,Polyakov:2015wna,Tseytlin:1990nb,
Siegel:1993bj,Hull:2004in,Condeescu:2012sp,Andriot:2012an,Wang:2015mia}
%

%%%%%%%%%%%%%%%%%%%%%%%%%%%%%
\subsubsection*{Branes, U-duality and M-theory}
%%%%%%%%%%%%%%%%%%%%%%%%%%%%%

\indent
Following \cite {Duff:1989tf}, it was pointed out  \cite{Duff:1989mp,Duff:1990hn,Duff:1990tb}  that 
membranes moving in a ($D\leq 4$)-dimensional space $M^D$ with coordinates $X^\mu(\tau,\sigma,\rho)$, 
background metric $g_{\mu\nu}(X)$ and 3-form $B_{\mu\nu\rho}(X)$ could usefully be described by a geometry 
with $[D+D(D-1)/2]$-dimensional coordinates
\be
Z^M=(X^{\mu},Y_{\rho\sigma})
\label{a}
\ee
and generalized metric
\be
G_{MN}= 
\left(
\begin{array}{cc}
g_{\mu\nu}+B_{\mu\rho\sigma}\,g^{\rho\sigma\lambda\tau}B_{\lambda\tau\nu}& 
B_{\mu\rho\sigma}\,g^{\rho\sigma\lambda\tau}\\
\,g^{\mu\nu\rho\sigma} \, B_{\rho\sigma\nu}&
g^{\mu\nu\rho\sigma}\\
\end{array}
\right)\,,
\label{b}
 \ee  
  
where
\be
g^{\alpha\beta\gamma\delta}=\frac{1}{2}(g^{\alpha\gamma}g^{\beta\delta} 
-g^{\alpha\delta}g^{\beta\gamma} )\,.
\ee

Once again, the motivation was twofold; worldvolume and spacetime:
\begin{enumerate}
\item{\bf Worldvolume}

In the case when $M^D$ is the $D$-torus $T^D$, the hope was to render 
manifest the M-theory U-dualities (using modern parlance) by combining 
worldvolume field equations and Bianchi identities. For example, the 
U-duality would be $SL(5,R)$ in the case $D=4$. The restriction to 
$D\leq4$ arises because, just as the usual coordinates $X^\mu$ correspond 
to momentum in the supersymmetry algebra, so the extra coordinates 
$Y_{\mu\nu}$ correspond to the M2 central charge. But for $D \geq 5$, 
this is not enough, as shown in Table 2 in \cite{Duff:1990hn}. There is 
also the M5 central charge with corresponding coordinates 
$Y_{\mu\nu\rho\sigma\tau}$, which first appears in $D=5$. 
In Appendix A, we illustrate the emergence of extra coordinates from 
central charges in the M-theory algebra for general $D$. For example, 
in the $D=7$ case  $X^\mu$, $Y_{\mu\nu}$, 
${\tilde Y}^{\mu\nu}\sim \epsilon^{\mu\nu\rho\sigma\tau\lambda\kappa}
Y_{\rho\sigma\tau\lambda\kappa}$ 
and $\tilde X^\mu$ form a 56 of the U-duality symmetry $E_{7(7)}$. 

\item{\bf Spacetime}

If this programme were successful, one would expect the 
$SL(D)/SO(D)$ of general relativity to be promoted not merely to  
$O(D,D) /(SO(D) \times SO(D))$ but to $E_8/SO(16)$, with possible 
infinite-dimensional extensions involving $E_9$, $E_{10}$ and $E_{11}$ 
as conjectured in \cite{Duff:1985bv,deWit:1986mz,West:2001as}. Once again, 
however, the generalised diffeomorphisms, section conditions and 
U-invariant actions came later. This activity has become known as 
``Exceptional Field Theory.''  For subsequent developments and variations 
on generalized geometry in M-theory and U-duality see, for example, 
\cite{Hull:2007zu,Pacheco:2008ps,Berman:2010is,Berman:2011pe,Berman:2011cg,
Berman:2011jh,Coimbra:2011ky,West:2012qz,Hatsuda:2012vm,Coimbra:2012af,
Park:2013gaj,Hatsuda:2013dya,Baguet:2015xha,Hohm:2013vpa,Hohm:2013uia,
Hohm:2014fxa,Berman:2013eva,deser,Hatsuda:2014aza,Blair:2014zba,Koepsell:2000xg}, 
where the 5-brane and other extended objects were incorporated, as required for $D>4$. 
The $E_{11}$ approach \cite{West:2001as} goes further with infinitely many coordinates 
of which those associated with the  M-theory central charges are but a subset.
	  
In summary, in contrast with strings where both the worldsheet and 
spacetime approaches have been successful, the brane worldvolume 
approach seems problematical and, with the exception of 
\cite{Berman:2010is,Hatsuda:2012vm}, recent developments have tended to 
focus on the spacetime approach where the extra coordinates (\ref{a}) and 
generalised metric (\ref{b}) have proved valuable. 
 In fact, the worldvolume approach has been questioned by Percacci and 
 Sezgin \cite{Percacci:1994aa},  by Sen \cite{Sen:1995cf}, and by 
Lukas and Ovrut \cite{Lukas:1997jk}. They suggest that it works only for 
target space dimensions $D=p+1$. In this case, the $D!/((D-p)! p!)$ wrapping
modes on a $D$-torus ($D\geq p+1$) and the $D$ Kaluza-Klein modes are
equal in number as in the case of a string. Sen argues that this equality is a 
requirement. If so, the $D=3$  U-duality $SL(3)\times SL(2)$ might be expected, 
but the $D=4$ U-duality $SL(5)$, would not.

In any event, the need to include coordinates corresponding to central 
charges in the M-theory algebra exposes a major difference between 
U-duality in M-theory and T-duality in string theory. In string theory, 
T-duality takes strings into strings, but in M-theory U-duality mixes up 
$p$-branes with different $p$. It seems unlikely, therefore, that the 
M2-brane worldvolume alone is sufficient. Somehow the totality of $p$-brane 
worldvolumes must conspire to give the full U-duality. This remains an 
unsolved problem.

Finally we note that the purpose of extra coordinates in both string and 
M-theory is to render the T and U dualities manifest. If one is content 
with non-manifest T-duality, one may invoke the Gaillard-Zumino (GZ) 
approach \cite{Gaillard:1981rj}, as was done in \cite{Cecotti:1988zz}. 
The GZ approach to U duality is discussed below.
 \end{enumerate}

\subsection{This Paper}

This paper is devoted purely to the worldvolume approach. We shall show:
\begin{itemize}
\item
There  is a problem with $SL(5,R)$, which manifests itself both
in the GZ approach (which doesn't introduce extra coordinates), 
as well as in approaches in which extra coordinates are introduced 
\cite{Duff:1990hn,Cederwall:2007je}. In the GZ approach, as well as the 
approach of \cite{Duff:1990hn}, 
we shall show that the obstacle to the realisation of $SL(5,R)$ symmetry is the 
nonintegrability of the transformation rule for the pull-back map. 
In the approach of \cite{Cederwall:2007je}, we shall show  that the proposed 
manifestly $SL(5,R)$-invariant equation for a membrane  in a target based 
on generalized geometry does not support linearized fluctuations about a 
Poincar\' e invariant vacuum solution.

\item 
In the case of topological membranes, the $SL(2,R)$ symmetry is known 
but we shall formulate it in  a double geometry setting.

\item We shall rederive the result that the membrane in $d+3$ dimensions 
has a Heisenberg subgroup of  the $SL(2,R)$ symmetry \cite{Percacci:1994aa}, 
by making use of simple integrability considerations.

\end{itemize}

%%%%%%%%%%%%%%%%%%%%%%
\section{Topological Membranes}
%%%%%%%%%%%%%%%%%%%%%%

Although $p$-branes in $p+1$ dimensions carry no dynamical degrees of 
freedom \cite{Fujikawa:1988hz}, they are nevertheless of considerable interest. 
In the present 
context, they provide us with a setting in which we can get a handle on 
duality symmetries that transform field equations and Bianchi identities 
into each other. Everything we will do here applies to topological 
$p$-branes for general $p$, but for simplicity in notation as well as our 
special interest in $M2$-branes, we shall focus on topological membranes.

The standard action for the closed membrane is
%%%%%
\be
I= \int d^3\sigma \Big[-\ft12  \sqrt{-\gamma}\, \gamma^{ij}
\del_i X^\mu \del_j X^\nu\, g_{\mu\nu}  +
   \ft16 \varepsilon^{ijk}\, \del_i X^\mu\, \del_j X^\nu\, \del_k X^\rho\,
B_{\mu\nu\rho}+\ft12\sqrt{-\gamma} \Big]\ ,
\label{sa}
\ee
%%%%%
where in our conventions $\varepsilon^{012}=+1$. For the topological membrane we take $\mu=0,1,2$\footnote{
This form of the action is in accordance with the terminology of `topological membrane' we are using here.  
It should be noted, however, that the characterization of membranes as `topological' also arises in the 
context of membranes that propagate in dimensions higher than three but with action that consists of 
Wess-Zumino term and no kinetic term. See, for example \cite{Ikeda:2006pd}.  In general, branes with 
pure Wess-Zumino terms exhibit a huge symmetry enhancement; see, for example \cite{Floreanini:1988hv}.} 
The $SL(3)$ is manifest.
For simplicity, we shall take the metric tensor $g_{\mu\nu}$ and
3-form potential $B_{\mu\nu\rho}$ in the 3-dimensional target space
to be constant\footnote{This is an important assumption. Otherwise the modification in 
equation (\ref{eom1})
will obstruct the sought after duality symmetry. 
The spacetime background here can be viewed
as being a subsector a time-like dimensional reduction of the bosonic sector of $D=11$ supergravity 
down to 8 dimensions, where only the fields  $(g_{\mu\nu}, B_{\mu\nu\rho})$ are kept, and the 8 dimensional 
Euclidean coordinates are taken to be constants. In a spacelike dimensional reduction, 
the signature of the metric  $g_{\mu\nu}$ would be Euclidean, and we would take the 8 dimensional 
spacetime coordinates to be constant. All of our considerations apply for this case as well.}.

In this case, using the algebraic field equation $\gamma_{ij}=\partial_i X^\mu \partial_j X^\nu\, g_{\mu\nu}$, 
we have the identity
\be
\sqrt{-\gamma} \gamma^{ij} g_{\mu\nu} \partial_j X^\nu =  
-\ft12 \sqrt{-g} \varepsilon_{\mu\nu\rho} \varepsilon^{ijk} \partial_j X^\nu \partial_k X^\rho\ .
\ee
Therefore, letting $B_{\mu\nu\rho}= \sqrt{-g} \varepsilon_{\mu\nu\rho} B$, the action can be written as
\be
I  = \frac1{3!} \int d^3\sigma\,  \sqrt{-g}\, (1+B)\, \varepsilon^{ijk}\, \del_i X^\mu\, 
\del_j X^\nu\, \del_k X^\rho\,\varepsilon_{\mu\nu\rho} \ .
\label{wza}
\ee
We shall, however, use the form \eq{sa} of the action below, motivated by the fact that this form 
will make it easier to compare with what happens in the case of the non-topological membrane.
The resulting equations are
%%%%%
\be
\del_i P^i_\mu = 0\ , \qquad
\gamma_{ij} =  \del_i X^\mu\, \del_j X^\nu\, g_{\mu\nu}
\;,
\label{eom1}
\ee
where
\be
\qquad P^i_\mu \equiv -\sqrt{-\gamma} \gamma^{ij}
\del_j X^\nu g_{\mu\nu} + \ft12 \varepsilon^{ijk}\, \del_j X^\nu \del_k X^\rho\, B_{\mu\nu\rho}\ .
\label{defP1}
\ee
%%%%%
There is also a conserved topological current:
%%%%%
\be
\del_i J^{i\mu\nu}=0\,,\qquad
J^{i\mu\nu}= \varepsilon^{ijk}\, \del_j X^\mu\, \del_k X^\nu\,.
\ee
%%%%%
Because we are considering a target space that is three dimensional, we
can make the definitions
%%%%%
\be
J^i_\mu=\ft12 \varepsilon_{\mu\nu\rho}\, J^{i\nu\rho}\,,\qquad
  B_{\mu\nu\rho} = B\, |g|^{1/2}\, \varepsilon_{\mu\nu\rho}\ .
\ee
%%%%%
Noting that
$\varepsilon^{ijk}\varepsilon_{\mu\nu\rho} \partial_j X^\nu \partial_k X^\rho
= -2 (\det \partial X) \gamma^{ij} g_{\mu\nu} \partial_j X^\nu$ and that
$\det \partial X = |\gamma|^{1/2} |g|^{-1/2}$, we can write $P^i_\mu$ and $J^i_\mu$
as follows
\be
P^i_\mu =-\sqrt{-\gamma} \gamma^{ij}g_{\mu\nu} (1+B)\, \del_j X^\nu\ ,\qquad
J^i_\mu = \sqrt{-\gamma} \gamma^{ij}g_{\mu\nu} |g|^{-1/2}\, \del_j X^\nu\ .
\ee
From these equations we find
\be
P^i_\mu = -(1+B)\, |g|^{1/2}\, J^i_\mu\ .
\label{PJ}
\ee
Note that this equation readily follows from the form of the  action given in \eq{wza}. 
We may now consider linear $GL(2,R)= SL(2,R) \times R$ 
transformations of the form
\footnote{It is understood that there also exists the trivial $SL(3,R)\times R^3$ symmetry realised as
$\delta P^i{}_\mu=\big(R_\mu{}^\nu
+S^{(\mu)}\delta_\mu^\nu\big) P^i{}_\nu$ and  $\delta
J^i{}_\mu=\big( R_\mu{}^\nu +S^{(\mu)}\delta_\mu^\nu
\big) J^i{}_\nu$,  where $R_\mu{}^\nu$ are real traceless matrices 
and $S^{(\mu)}$ are the real scaling parameters.}
%
%%%%%
\be
 \delta  \left(
        \begin{array}{c}
           P^i_\mu \\
           J^i_\mu \\
         \end{array}
       \right)
=   \left(
      \begin{array}{cc}
        a & b \\
        c & -a \\
      \end{array}
    \right)  \left(
         \begin{array}{c}
           P^i_\mu \\
           J^i_\mu \\
         \end{array}
       \right)
       + \lambda  \left(\begin{array}{c}
           P^i_\mu \\
           J^i_\mu \\
         \end{array}
       \right)\ .
\label{matrix}
\ee
%%%%%
We see that \eq{PJ} is left invariant provided that $|g|$ and $B$ transform such that
%%%%%
\be
C\equiv - (1+B)\, |g|^{1/2}
\label{defC}
\ee
%%%%%
transforms as
%%%%%
\be
\delta C = b + 2a C - c\, C^2\ .
\label{dC}
\ee
%%%%%
This can be seen from (\ref{PJ}), noting that it implies  
$C\oneone = P J^{-1}$. It represents the infinitesimal
form of a fractional linear transformation of the real variable $C$, and 
gives a representation
of the algebra $SL(2,R)\times R$. The fact that this symmetry acts on 
a combination of $g$ and $B$ is a
consequence of the fact that the target spacetime is Lorentzian, 
as noted in \cite{Percacci:1994aa}.

To make the $SL(2,R)$ symmetry manifest, we introduce a doubled
system of coordinates $Z^{a\mu}$, with $a=1,2$, such that 
$X^\mu=-Z^{1\mu}$ and
\bea
P^i_\mu = \sqrt{-\gamma} \gamma^{ij}  |g|^{-1/2}\,g_{\mu\nu}\,  \partial_j Z^{2\nu}  \ ,
\qquad  J^i_\mu =    -\sqrt{-\gamma} \gamma^{ij}  |g|^{-1/2}\,g_{\mu\nu}\, \partial_j Z^{1\nu}\ .
\label{PJR}
\eea
This doubling of coordinates is in accordance with the generalized 
target space geometry
recently studied in \cite{Hohm:2015xna} for maximal supergravity in 
eight dimensions.
Using these definitions, it follows that \eq{PJ} can be written as
\bea
\partial_i Z^{a\mu} &=& G^{ab}\epsilon_{bc}\, \partial_i Z^{c\mu}\,,
\label{covariant}
\eea
where
\bea
G^{ab} &=&
\left(
\begin{array}{cc}
|g|^{-1/2} &\quad B\\
B& \quad |g|^{1/2}\,
(B^2-1)
\end{array}
\right)
\ ,
\eea
transforming by conjugation under ${\rm SL}(2,R)$. 
Note that ${\rm det}\,G_{ab}=-1$, such that the product
$({\rm det}\,G_{ab})({\rm det}\,g_{\mu\nu})=1$.
Denoting the inverse of this metric by $G_{ab}$, it transforms under 
infinitesimal $SL(2,R)$ transformations as  
$\delta G_{ab}= \Lambda_a{}^c G_{cb} + \Lambda_b{}^c G_{ac}$, with $\Lambda$ as 
given in \eq{matrix}. Written out, this  gives
\be
\delta g^{1/2} = 2 c\, |g| B -2d\, |g|^{1/2}\ ,
\qquad
\delta B = -b\, |g|^{-1/2} - c\, |g|^{1/2} (B^2-1)\ .
\label{deltagB}
\ee
Using these rules, the transformation of $-|g|^{1/2} (1+B)$ indeed gives 
the result \eq{dC}. 
Demanding manifest $SL(2,R)$ invariance thus fixes the separate variation of $g$ and $B$ under 
$SL(2,R)$, not just its combination (\ref{defC}).

It is important to note that the $SL(2,R)$ transformation under which $(\partial_i Z_{1\mu},\partial_i Z_{2\mu})$ 
forms a doublet is embedded
into \eq{matrix} with a field-dependent scale transformation with parameter
\be
\lambda= -a + \frac13 c (3+B)\ ,
\ee
as can be determined from \eq{PJR}.

Turning to the key equation \eq{covariant}, it can be written as 
$P_{ab} \partial_i Z^{b\mu}=0$, where $P_{ab} = 
\ft12(G_{ab}-\epsilon_{ab})$ is a projector, and it amounts to
\be
\partial_i Z^{2\mu} = |g|^{1/2} (1+B) \partial_i Z^{1\mu}\ .
\label{z12}
\ee
Acting with $\nabla_i$ does not yield the field equation 
$\partial_i \left(\sqrt{-\gamma} \gamma^{ij} g_{\mu\nu} 
\partial_j X^\nu\right)=0$.
However, the latter is identically satisfied for the topological 
membrane, since 
\be
\partial_i \left(\sqrt{-\gamma} \gamma^{ij} g_{\mu\nu} \partial_j X^\nu\right) \equiv  
-\ft12 \partial_i \left(\sqrt{-g} \varepsilon_{\mu\nu\rho} \varepsilon^{ijk} 
\partial_j X^\nu \partial_k X^\rho\right)=0\ ,
\ee
recalling that the target space metric is constant. 

It is instructive to perform a double dimensional reduction 
\cite{Duff:1987bx} of \eq{covariant}.  To this end, we let
\be
{\hat g}_{\hat\mu\hat\nu} =  \left(
\begin{array}{cc}
\phi\, g_{\mu\nu} &0 \\
0& \phi^{-2}
\end{array}
\right)\ ,\qquad {\hat B}_{\mu\nu 2} = |g|^{1/2} \varepsilon_{\mu\nu}\, B\ ,
\ee
where $g_{\mu\nu}, \phi$ and $B$ are constants.  Note that $\sqrt{-\hat g} = \sqrt{-g}$. 
Choosing the gauge $X^2=\sigma^2$ and letting $\partial_2 Z^{a\mu}=0$ gives
\be
{\hat\gamma}_{\hat i\hat j} =  \left(
\begin{array}{cc}
\gamma_{ij}&0 \\
0& \phi^{-2}
\end{array}
\right)\ , \qquad     \partial_i X^\mu \partial_j X^\nu g_{\mu\nu} 
-\frac12 \gamma_{ij} \gamma^{k\ell} \partial_k X^\mu \partial_\ell X^\nu g_{\mu\nu}=0\ .
\ee
It follows that \eq{covariant}, or equivalently \eq{z12}, holds for the 
topological string, where the indices now run over two values, namely, 
$i=0,1$ and $\mu=0,1$. Setting $\hat i=2$ and $\hat\mu=2$ in  \eq{z12}  
fixes $Z^{22}$, giving it a linear dependence on the coordinate $\sigma^2$. 
Comparing \eq{covariant} for the topological string with \eq{dcon}, they are
in fact,
contrary to appearance, the same equation, with the identification 
$x^\mu = -Z^{1\mu}$ and $y_\mu = -\varepsilon_{\mu\nu} Z^{2\nu}$.  
This can be seen by writing \eq{dcon} as
\bea
G^{ab} \varepsilon_{bc} \varepsilon_{\mu\nu} \varepsilon^{ij}\partial_{j}Z^{c\nu}=
 \bar{g}_{\mu\nu} \sqrt{-\gamma}\gamma^{ij}\partial_{j}Z^{a\nu}\ ,
\label{1}
\eea
where ${\bar g}_{\mu\nu} = |g|^{-1/2} g_{\mu\nu} $. Thus
\bea
G^{1b} \varepsilon_{bc} \varepsilon_{\mu\nu} \varepsilon^{ij}\partial_{j}Z^{c\nu}=
 \bar{g}_{\mu\nu} \sqrt{-\gamma}\gamma^{ij}\partial_{j}Z^{1\nu}
 =-\varepsilon^{ij} \varepsilon_{\mu\nu} \partial_{j}Z^{1\nu}\ ,
\eea
where we have used the formula for the determinant of\ $\partial_j Z^{1\nu}$\  in the second equation.
From this we conclude 
\bea
G^{1b} \varepsilon_{bc}\partial_{j}Z^{c\nu}=
- \partial_{j}Z^{1\nu}\ ,
\eea
showing the equivalence of \eq{covariant} and \eq{dcon}, up to a 
relative sign which can be attributed to convention choices.

%%%%%%%%%%%%%%%%%%%%%%%%%%%%%%%%%
\section{Membrane in $D=8$}
%%%%%%%%%%%%%%%%%%%%%%%%%%%%%%%%%

%\textcolor{blue}{
The M2-brane action in $D=8$ can be obtained from the M2-brane action in $D=11$ by dimensional 
reduction on 3-torus.  The bosonic sector of such a reduction has been studied in \cite{Lukas:1997jk} where 
$SL(3,R)\times SL(2,R)$ symmetry could not be 
established. In a different approach aiming at a direct construction of an M2-brane action in $D=8$ which 
couples to all  the bosonic fields of the maximal supergravity theory in which the $SL(3,R)\times SL(2,R)$
 symmetry is built in manifestly has been proposed \cite{Bengtsson:2004nj}. However, the condition of $SL(2,R)$ 
 symmetry puts nonlinear constraints on the field which have been solved only in a fashion that exhibits a two 
 parameter subgroup of $SL(2,R)$ as a symmetry. More specifically, the bosonic sector of maximal $D=8$ 
 supergravity has the fields
%}
%
%\textcolor{blue}{
\be
(g_{\mu\nu}, B_{\mu\nu\rho}, C_{\mu\nu m}, A_\mu^{mr}, 7 \phi)\ , \qquad m=1,2,3,\quad r=1,2
\ee
%}
%
%\textcolor{blue}{
where the seven scalars parametrize the coset  $(SL(3,R)/SO(3))\times (SL(2,R)/SO(2))$, the vector 
fields transform as (3,2) of $SL(3,R)\times SL(2,R)$. The field strength of the 3-form field is combined 
with the dual field strength for a doublet of $SL(2,R)$.  The gauge invariance of the pullbacks of the 
field strengths of the $1,2,3$-form potentials requires the introduction of worldvolume fundamental 
potentials, resulting in the field strengths 
$h^r= db^r- \underline{B}^r+\cdots, \ g_m= dc_m -\underline{C}_m +\cdots$ and 
$f^{mr}=d\phi^{mr}-\underline A^{mr}$, where the underlined fields are the pullback of the target space forms. 
The action proposed in \cite{Bengtsson:2004nj} then takes the form 
$I=\int \sqrt{-\gamma} \lambda(1+\Phi(f,g) + \star h^r \star h^s G_{rs})$ where $\lambda$ is a Lagrangian 
multiplier field, $G_{rs}$ is $SL(2,R)$ matrix parametrized in terms of the $SL(2,R)/SO(2)$ coset scalars 
and $\Phi$ is a function of the field strengths $(f^{mr},g_m)$. Duality relations for these field strengths are 
imposed by hand in addition to the field equations that follow from the action to ensure the correct number 
of propagation degrees of freedom, namely the 5 scalars coming from $X^\mu$ and 3  scalars  $\phi^{m1}$. 
The resulting field equations have not lent themselves to a solution in general, however, and a special 
solution discussed in \cite{Bengtsson:2004nj} breaks $SL(2,R)$ symmetry.
%}

%\textcolor{blue}{
Our approach here is instead to consider a membrane propagating in $D=8$ dimensions and coupled to the 
target space metric and 3-form potential only, and to study the duality symmetry of the standard membrane action. 
The background can be viewed as the truncated version of the maximal supergravity.  In fact, all considerations 
below apply equally well to $p$-branes in  $d+p$ dimensions propagating in the background of a metric and 
$p+1$ potential. We thus consider an $8+3$ dimensional space-time with coordinates
%}

%
\bea
X^{\hat \mu} = (x^\mu, y^\alpha)\;,\qquad
\mu=0, \dots, 7\;,\quad \alpha=1,2, 3\;,
\eea
with $x^0$ being in the time direction. 
We take the background geometry to have the form
\bea
g_{\hat\mu\hat\nu} &=& 
\left(
\begin{array}{cc}
{g}_{\mu\nu}(x) &0\\0& g_{\alpha\beta}(x) 
\end{array}
\right)
~=~
\left(
\begin{array}{cc}
g_{(3)}^\kappa\,\bar{g}_{\mu\nu}&0\\0&g_{(3)}^{1/3}\,\bar{g}_{\alpha\beta}
\end{array}
\right)
\;,
\eea
where $g_{(3)}\equiv {\rm det}\,g_{\alpha\beta}$ and
$\bar{g}_{\mu\nu}$, $\bar{g}_{\alpha\beta}$ are assumed to be
$SL(2,R)$ invariant, and $\kappa$ is an exponent to be determined. 
We take the only nonvanishing component of the 3-form to be $B_{\alpha\beta\gamma}$,
and assume that all the target space background fields to depend on $x^\mu$ only. 
%
%\textcolor{blue}{
Thus, the action is 
\bea
I &=&\int d^3\sigma \Bigl[ -{1\over2}\sqrt{-\gamma}
\gamma^{ij} \left(\partial_i x^\mu\partial_j x^\nu g_{\mu\nu}(x)
+\partial_iy^\alpha \partial_j y^\beta g_{\alpha\beta}(x) \right)
+\frac12 \sqrt{-\gamma}
\nn\\
&& \qquad\quad  +\frac1{3!}  \varepsilon^{ijk} \partial_i y^\alpha \partial_j y^\beta \partial_k y^\gamma
B_{\alpha\beta\gamma} (x) \Bigr]\ .
\label{8da}
\eea
In maximal supergravity theory in $D=8$, the fields $(g_{\alpha\beta}, B_{\alpha\beta\gamma})$ 
contain five scalars that parametrise the coset $SL(3,R)/SO(3)$ and two scalar parametrising the 
coset $SL(2,R)/SO(2)$. In addition to these fields and the metric $g_{\mu\nu}$ there are two triplet 
of vectors,  $g_{\mu\alpha}$ and $B_{\mu\alpha\beta}$, that transform as (3,2) under three $2$-forms 
$B_{\mu\nu\alpha}$ that transform as $(3,1)$ under the U-duality group $SL(3,R) \times SL(2,R)$. 
We are neglecting the latter fields below, with the expectation that they would not effect the realisation 
of $SL(3,R)\times SL(2,R)$ U-duality symmetry at the level of duality rotations on the worldvolume of 
the membrane in $D=8$, should such symmetry exist at all\footnote{All fields have been kept in \cite{Lukas:1997jk} 
where $SL(3,R)\times SL(2,R)$ duality as duality rotation symmetry
on the membrane worldvolume is sought but not found.}.
%}

Turning the action \eq{8da}, it implies that the induced worldvolume metric is given by
\bea
\gamma_{ij} &=&  g_{(3)}^\kappa\,\bar{g}_{\mu\nu} \,\partial_i x^\mu \partial_j x^\nu
+ g_{(3)}^{1/3}\,\bar{g}_{\alpha\beta}\,\partial_i y^\alpha \partial_j y^\beta
\;,
\eea
and the field equations are
%%%%%
\be
\del_i P^i_\mu = S_\mu\ , \qquad
\del_i P^i_\alpha = 0
\;,
\label{eomPP}
\ee
where
\bea
\qquad P^i_\alpha &\equiv& -\sqrt{-\gamma} \gamma^{ij}
\del_j y^\beta g_{\alpha\beta} + \ft12 \varepsilon^{ijk}\, \del_j y^\beta \del_k y^\gamma\, B_{\alpha\beta\gamma}\;,
\nonumber\\
\qquad P^i_\mu &\equiv& -\sqrt{-\gamma} \gamma^{ij}
\del_j x^\nu g_{\mu\nu} \ ,
\nonumber\\
S_\mu & \equiv& -\frac12 \sqrt{-\gamma}\gamma^{ij} \partial_i x^\nu \partial_j x^\rho \partial_\mu g_{\nu\rho}
-\frac16 \varepsilon^{ijk} \partial_i y^\alpha \partial_j y^\beta \partial_k y^\gamma\, \partial_\mu B_{\alpha\beta\gamma}\ .
\eea
%%%%%
There is also a conserved topological current:
%%%%%
\be
\del_i J^{i\alpha\beta}=0\,,\qquad
J^{i\alpha\beta}\equiv \varepsilon^{ijk}\, \del_j y^\alpha\, \del_k y^\beta
\label{JJ}
\;,
\ee
%%%%%
such that, defining
\bea
B_{\alpha\beta\gamma} (x) &=& \varepsilon_{\alpha\beta\gamma}\,g_{(3)}^{1/2}\,B\ ,
\eea
we have
\bea
 P^i_\alpha &=& -\left(
 \sqrt{-\gamma} \gamma^{ij}
+ \sqrt{V} V^{ij}\, B \right) g_{\alpha\beta}\,\del_j y^\beta \;,
\nonumber\\
J^i_\alpha&\equiv&\frac12 \varepsilon_{\alpha\beta\gamma}\, J^{i\beta\gamma}
~=~ \sqrt{V} V^{ij}\, g_{(3)}^{-1/2}\, g_{\alpha\beta} \del_j y^\beta\;,
\label{PJ83}
\eea
where $V^{ij}$ is the inverse of
\bea
V_{ij}\equiv \partial_i y^\alpha \partial_j y^\beta\,{g}_{\alpha\beta}
\;.
\eea
We can combine these equations as
\be
P^i_\alpha = -|g|^{1/2} (\delta^i_j + B X^i{}_j ) J^j_\alpha\ ,
\label{newPJ}
\ee
where
%%%%%
\be
   X^i{}_j  = \fft{(\gamma^{-1}\, V)^i{}_j}{\sqrt{-\det (\gamma^{-1}\, V)}}\,.
\ee
%%%%%

In \cite{Percacci:1994aa}, the self-consistency of \eq{newPJ} was studied 
in detail, and it was shown that a two-parameter subgroup of $SL(2,R)$ 
can be realized. In doing so, the complicated transformation rule for the induced metric, 
which is no longer a scaling transformation we saw in the case of topological membrane, 
was taken into account in\cite{Percacci:1994aa}\footnote{In \cite{Duff:1990hn} only the 
second term is kept, and therefore it effectively deals with membrane in
$D$ dimensional target where $\alpha=1,...D$.}.

Here, we shall avoid this complication and  show that this same result can be derived more 
simply from the integrability of the transformation rule 
for $\partial_i y^\alpha$. Assuming that $P_\alpha^i$ and $J_\alpha^i$ 
transform as in \eq{matrix}, we can compute the transformation of
$\partial_i y^\alpha$ by using (\ref{JJ}) and
(\ref{matrix}), finding 
%%%%%
\be
\delta\del_i y^\alpha = \left[ \left( -a  +c (1+ \frac13 B +\frac12 {\rm tr} X) \right) \delta^j_i
          -c X^j{}_i \right] \del_j y^\alpha\ .\label{cdtrans}
\ee
%%%%%

As discussed above, the integrability of this variation is in general 
not guaranteed, even on-shell.
Clearly the $a$ transformation is always integrable, and so the remaining
question is whether the $c$ transformation in (\ref{cdtrans}) is integrable.
In order to test this on-shell, it will suffice to consider a 
particular membrane solution \cite{Hoppe:1987vv},  for which 
$g_{\hat\mu\hat\nu}=\eta_{\hat\mu\hat\nu}$ and we take 
\bea
x^0 &=&\sigma^0\,,\qquad x^\mu=\hbox{constant}\ \hbox{for}\ 1\le\mu\le 7\,,
\nn\\
y^1&=& \alpha\, \sigma_1\, \cos(\omega \sigma^0)\,,\qquad
y^2= \alpha\, \sigma_1\, \sin(\omega\sigma^0)\,,\qquad
y^3=\beta\, \sigma_2\,.\label{xyback}
\eea
%%%%%
Thus we have the induced metric 
%%%%%
\be
\gamma_{ij}= \del_i y^\alpha\, \del_j y^\beta \, \delta_{\alpha\beta} + 
   \del_i x^\mu\, \del_j x^\nu\, \eta_{\mu\nu}\,,
\ee
%%%%%
giving
%%%%%
\be
\gamma_{00}= -1+ \alpha^2 \omega^2 \sigma_1^2\,,\qquad\gamma_{11}=\alpha^2
\,,\qquad
     \gamma_{22}= \beta^2\,.
\ee
It is straightforward to verify that this is a solution of the
equations of motion.
   We see that in this background
%%%%%
\bea
X-\ft12 {\rm tr}\, X &=& {\rm diag}\, (f,h,h)\,,\nn\\
f&=& -\ft12\alpha\omega\sigma_1\, (1-\alpha^2\omega^2\sigma_1^2)^{-1/2} -
    (\alpha\omega\sigma_1)^{-1}\, (1-\alpha^2\omega^2\sigma_1^2)^{1/2}\,,
\nn\\
h&=& \ft12\alpha\omega\sigma_1\, (1-\alpha^2\omega^2\sigma_1^2)^{-1/2}\,.
\label{fhdef}
\eea
%%%%%
Let us consider the transformations 
$\delta\del_1 y^2$ and $\delta\del_2 y^2$ under $c$, which from 
(\ref{xyback}), (\ref{cdtrans}) and (\ref{fhdef}) will therefore give
%%%%%
\be
\delta \del_1 y^2=0\,,\qquad \delta\del_2 y^2= c\beta\, h\,.
\ee
%%%%%
Checking the integrability, we see from the first equation that
$\del_2\delta\del_1 y^2=0$, whereas from the second equation 
$\del_1\delta\del_2 y^2 = c\beta \ft{\del h}{\del\sigma_1}\ne0$.  This
example is therefore sufficient to show that the proposed transformation
for $\delta\del_i y^\alpha$ under the $c$ transformation, subject only
to the use of the membrane equations of motion,  is not integrable.

Choosing $\kappa=-\frac1{6}$, equations (\ref{newPJ}) transform properly 
under the remaining $SL(2,R)$ transformations
\bea
\left(
\begin{array}{cr}
a&b\\0&-a
\end{array}
\right)\;,
\label{ha}
\eea
provided the background parameters $g$ and $B$ transform according to (\ref{deltagB}).
As a consequence, the transformation of the induced metric takes the form
\bea
\delta \gamma_{ij} &=& -\frac{2a}{3}\,\gamma_{ij}\;,\qquad
\delta V_{ij} = -\frac{2a}{3}\,V_{ij}
\eea
Equations \eq{PJ} can be combined as 
\be
X^j{}_i \,\partial_j Z^{a\alpha} = G^{ab} \epsilon_{bc} \partial_i Z^{c\alpha}\ .
\ee
The presence of the matrix $X^i{}_j$ shows that the duality symmetry of this equation is the Heisenberg group 
with the underlying algebra parametrized as in \eq{ha}.

%%%%%%%%%%%%%%%%%%%%%%%
\section{Membrane in $D=4$}
%%%%%%%%%%%%%%%%%%%%%%%

In this section we shall study the action \eq{sa} for $D=4$ target spacetime with Lorentzian signature. 
In this action \eq{sa}, the target space fields can be interpreted as  the bosonic sector of $N=1, D=4$ supergravity 
in which the cosmological constant is dualised to a 3-form potential \cite{Gates:1980ay,Binetruy:1996xw,Ovrut:1997ur}.  
Supermembrane action  in this setting exists \cite{Bergshoeff:1987cm,Achucarro:1987nc} and it has been  studied in detail 
in \cite{Ovrut:1997ur}. The spacetime background can also be viewed as being a subsector a time-like dimensional 
reduction of the bosonic sector of $D=11$ supergravity down to 7 dimensions, where only the fields  
$(g_{\mu\nu}, B_{\mu\nu\rho})$ are kept, and the 7 dimensional Euclidean coordinates are taken to be constants. 
In a spacelike dimensional reduction,  the signature of the metric  $g_{\mu\nu}$ would be Euclidean, and again, 
we would take the 7 dimensional spacetime coordinates to be constant. Our considerations apply 
to this case as well but we shall adhere to the Lorentzian signature for concreteness.

We shall consider two approaches to the problem of duality rotations in 
this theory.
In the first approach, due to Gaillard and Zumino 
\cite{Gaillard:1981rj}, there is no need to introduce any extra 
coordinates.  Rather, one examines the consistency of the duality 
rotations, since the definition of the conjugate momentum field 
associated with the worldvolume scalar fields involves the 
topological current whose conservation is the Bianchi identity. 
In a second approach considered in \cite{Duff:1989mp,Duff:1990hn,Duff:1990tb}, 
one introduces extra coordinates in order to try to achieve 
a manifest realization of the duality symmetry.

%%%%%%%%%%%%%%%%%%%
\subsection{Gaillard-Zumino approach}
%%%%%%%%%%%%%%%%%%%

The equations of motion for a membrane in a four-dimensional target space 
with coordinates $X^\mu$,  $(\mu=0,...,3)$  can be written as
\be
\del_i P^i_\mu = 0\ , \qquad
\gamma_{ij} =  \del_i X^\mu\, \del_j X^\nu\, g_{\mu\nu}\,,
\label{eom2}
\ee
where
\be
\qquad P^i_\mu \equiv  - 
    g_{\mu\nu} {\cal F}^{i\nu} + \ft12  B_{\mu\nu\rho} J^{i \nu\rho}\ 
\label{defP}
\ee
and
\be
{\cal F}^{i\mu} \equiv \sqrt {-\gamma} \gamma^{ij} \partial_j X^\mu\,\qquad 
J^{i\mu\nu} \equiv \varepsilon^{ijk} \partial_j X^\mu \partial_k X^\nu\ .
\label{defJP}
\ee
It is important to note that \eq{defJP} implies the 
relation\footnote{We use the convention $\varepsilon^{012}=
-\varepsilon_{012}=1$}
\be
%- \gamma 
J^{i\mu\nu} = 
\gamma^{i\ell} \varepsilon_{\ell jk} {\cal F}^{j\mu} {\cal F}^{k\nu}\ .
\label{CR}
\ee
It will also be useful to note the relation
\be
-\gamma \gamma^{ij} = {\cal F}^{i\mu} {\cal F}^{j\nu}\, g_{\mu\nu} \ .
\label{FF}
\ee
The question then is what is the largest set of  transformations 
that transforms $P^i_\mu$ and $J^{i\mu\nu}$ into each
other in a consistent manner, such that the system of field equations 
$\partial_i P^i_\mu=0$ and the Bianchi identity 
$\partial_i J^{i\mu\nu}=0$ remain invariant. 

Our task is to check whether the equations of motion are invariant 
under $SL(5,R)$ transformations 
which can be parametrised as
\be
\Lambda^M{}_N = \left(
\begin{array}{cc}
a^\mu{}_\nu +\frac14 a\, \delta^\mu_\nu  & \qquad  
-\ft16\varepsilon^{\mu\nu\rho\sigma} b_{\nu\rho\sigma} \\
\ft16 
\varepsilon_{\nu\mu\rho\sigma}\, c^{\mu\rho\sigma} & \qquad -a
\end{array}
\right)\ ,
\ee
where $a^\mu{}_\nu$ is traceless. It acts on a $5$-plet of $SL(5,R)$ as 
$\delta V_M = -\Lambda^P{}_M\,V_P$. Thus,
defining the components of a triplet of second-rank antisymmetric tensor of 
$K^i_{MN}$ as 
$K^i_{\mu 5} := P^i_{\mu}$ and $K^i_{\mu\nu} := \ft12 
\varepsilon_{\mu\nu\rho\sigma}\, J^{i\rho\sigma}$,  it follows from 
$\delta K^i_{MN} = 2 \Lambda^P{}_{[M}\,K^i_{N]P}$ that
 \bea
   \delta  \left(
        \begin{array}{c}
           P^i_\mu \\
           J^{i\,\mu\nu} \\
         \end{array}
       \right)
&=& 
\left(
\begin{array}{cc}
-a^\rho{}_\mu +\frac34 a\delta^\rho_\mu & \qquad \frac12 b_{\mu\rho\sigma}\\
c^{\mu\nu\rho} & \qquad 2 a^{[\mu}{}_{[\rho} \delta^{\nu]}_{\sigma]}
-\frac12 a \delta^{\mu\nu}_{\rho\sigma} 
\end{array}
\right)\,
\left(
        \begin{array}{c}
           P^i_\rho \\
           J^{i\,\rho\sigma} \\
         \end{array}
       \right)
        \label{t12}
\label{dPJ}
 \eea
where $\delta^{\mu\nu}_{\rho\sigma}=\frac{1}{2}
\left(\delta^\mu_\rho\delta^\nu_\sigma
 -\delta^\nu_\rho\delta^\mu_\sigma\right)$. 
 Next, assembling $(g_{\mu\nu}, B_{\mu\nu\rho})$ into a symmetric $SL(5,R)$ 
matrix $G_{MN}$, where $M,N=1,...,5$, with identifications
\bea
G_{\mu\nu} &=&  g^{-2/5} g_{\mu\nu}\ , \quad 
G_{\mu 5}=G_{5\mu} 
= \frac{1}{3!} g^{-2/5} 
g_{\mu\alpha} \varepsilon^{\alpha\beta\gamma\delta} B_{\beta\gamma\delta}\ ,
\nn\\
G_{55} &=& g^{3/5} \left(1+ \frac{1}{3!} B^2\right)\ ,
\qquad B^2 \equiv B^{\mu\nu\rho} B_{\mu\nu\rho}\ ,
\eea
it follows from  $\delta G_{MN} = -2\Lambda^P{}_{(M}\,G_{N)P}$ that
\bea 
 \delta g_{\mu\nu} &=& -2 a^\sigma{}_{(\mu} \,g_{\nu)\sigma} 
 +\frac56 \left( a 
 +\frac{2}{15} c\cdot B \right)  g_{\mu\nu}
 -c^{\alpha\beta}{}_{(\mu} B_{\nu)\alpha\beta} \ ,
 \label{t3}\\
 \delta B_{\mu\nu\rho} &=& -3 a^\sigma{}_{[\mu} \, B_{\nu\rho]\sigma}  
  +\frac54 \left( a -\frac{2}{15} c\cdot B\right) B_{\mu\nu\rho} 
+b_{\mu\nu\rho} +  
\, c_{\mu\nu\rho}\ ,
 \label{t4}
\eea
where indices are lowered on the parameters $c^{\mu\nu\rho}$ using the
metric $g_{\mu\nu}$, and  we have defined 
$c\cdot B \equiv c^{\alpha\beta\gamma} B_{\alpha\beta\gamma}$.  
Next, the variation
of ${\cal F}^{i\mu}$ can be found from \eq{defP} by using the 
variations \eq{dPJ}, \eq{t3} and \eq{t4}. 
The result is
 \be
 \delta {\cal F}^{i\mu} = a^\mu{}_\nu {\cal F}^{i\nu} 
 -\frac1{12} \left( a+\frac43 c\cdot B\right) {\cal F}^{i\mu}
 +\frac12 c^\mu{}_{\nu\rho} J^{i\nu\rho}
 + \frac12 c^{\mu\nu\rho} B_{\nu\rho\sigma} {\cal F}^{i \sigma}\ .
 \label{t5}
 \ee
In deriving this variation,  one makes use of the identity
$B^\lambda{}_{\alpha\beta}c^{\alpha\beta\sigma}
 B_{\sigma\nu\rho}
=\frac{1}{3}c\cdot B B^\lambda{}_{\nu\rho}$,
which can be proven by writing $c^{\mu\nu\rho}$ and
$B_{\mu\nu\rho}$ in terms of dual vector fields.
Finally, from the variations above, one can also determine the variation of 
$\gamma_{ij}$ by using \eq{FF}, finding
\be
\delta\gamma_{ij} = 
\frac13 \left(a -\frac1{3!} c\cdot B +\frac12 \varepsilon\right) \gamma_{ij}\ ,
\label{vargamma}
\ee
where
\be
\varepsilon \equiv  \fft1{\sqrt{-\gamma}}\,\varepsilon^{ijk} 
\partial_i  X^\mu \partial_j X^\nu \partial_k X^\rho\, c_{\mu\nu\rho} \ .
\ee

Now we turn to a key test for the above transformation rules, which is the requirement that 
the transformations of  $J^{i\,\mu\nu}$, ${\cal F}^{i\,\mu}$, and 
$\gamma_{ij}$, must be compatible with equation \eq{CR}.  The question 
of whether this nontrivial condition holds was raised in 
\cite{Percacci:1994aa}. In fact, as we shall show here, it does 
actually hold. Firstly, the invariance of \eq{CR} under the $a$- and 
$b$-dependent transformations is manifest. The nontrivial check is the 
invariance under the $c$-dependent transformations, which requires that 
\bea
0 &=& \left( J_i^{\sigma[\mu} c^{\nu]\alpha\beta} + 
\frac12  J_i^{\alpha\beta} c^{\mu\nu\sigma} 
+\frac16 J_i^{\mu\nu} c^{\alpha\beta\sigma} \right) B_{\alpha\beta\sigma} 
\nn\\
&& {-\varepsilon_{ijk} J^{j\alpha\beta} {\cal F}^{k[\mu} c^{\nu]}{}_{\alpha\beta} }
+\frac16 \varepsilon J_i^{\mu\nu} - c^{\mu\nu}{}_\sigma {\cal F}_i^\sigma\ .
\eea
The first three terms sum up to zero since 
$J_i^{[\sigma\mu} c^{\nu\alpha\beta]} =0$ identically. 
The remaining three terms also sum up to zero,
upon using the fact that $\gamma^{ij}\partial_i X^\mu\partial_j X^\nu=
g^{\mu\nu}-n^\mu n^\nu$,
where $n^\mu$ is normal to the brane in the target space, i.e. $\del_i X^\mu\,
g_{\mu\nu}\, n^\nu=0$.
There remains, however, the condition that the variation of  
$\partial_i X^\mu$, which follows from the first equation in \eq{defJP}, 
must be curl-free.  Using \eq{t5} and (\ref{vargamma}) in the first 
equation in  \eq{defJP} we find that
\be
\delta \partial_i X^\mu = \left[ a^\mu{}_\sigma 
-\frac14 \left( a  
+\frac13 c\cdot B  
 +\frac13 \varepsilon \right) \delta^\mu_\sigma      + \frac12 c^{\mu\alpha\beta} 
 B_{\alpha\beta\sigma} \right]  \partial_i X^\sigma 
+ \frac12\ c^\mu{}_{\nu\rho} \frac{\gamma_{ij} J^{j\nu\rho}}{\sqrt{-\gamma}}\ .
\ee
Thus the integrability condition amounts to
\bea
0=\varepsilon^{ijk}\, \del_j\delta\del_k X^\mu &=&
-\ft1{12} \varepsilon^{ijk} \del_j\varepsilon\, \del_k X^\mu 
+ \ft12 \varepsilon^{ijk} c^\mu{}_{\rho\sigma}\, \del_j\Big(
   \fft{\gamma_{k\ell}\, J^{\ell\rho\sigma}}{\sqrt{-\gamma}}
\Big)\ .
\label{integrability}
\eea
Using the field equation  $\nabla_i \partial^i X^\mu =0$, this equation can be simplified to read
\be
0=V_k^{\rho\sigma} \left( \gamma^{ki} g^{\mu\nu} 
+ 2\gamma^{k[m} \gamma^{i]n} \partial_m X^\mu \partial_n X^\nu \right)  c_{\nu\rho\sigma} \ ,
\ee
where we have defined
\be
V_i^{\mu\nu}\equiv \nabla_i \partial^j X^{[\mu} \partial_j X^{\nu]}\ .
\label{anomaly}
\ee
As we did earlier, we can most conveniently check this  equation 
by considering a  particular membrane solution~\cite{Hoppe:1987vv}, namely
%%%%%
\be
X^\mu= (\sigma^0, \alpha \sigma_1\, \cos(\omega \sigma^0), 
     \alpha \sigma_1\, \sin(\omega \sigma^0), \beta \sigma_2)\,,
\ee
%%%%%
which solves the equations of motion (\ref{eom2}).
If suffices to consider the integrability
condition (\ref{integrability}) for $i=0$ and $\mu=0$. 
It is then immediately 
evident that the first term on the right-hand side of (\ref{integrability})
gives zero, whereas the second term gives a non-vanishing result that is
proportional to the parameter $c^0{}_{12}$.
Thus the integrability condition is not 
satisfied, and so the proposed $SL(5,R)$ transformation of 
$\del_i X^\mu$
is not valid.

The subgroup that is consistent with the curl-free condition is 
therefore the semi-direct
product ${\rm GL}(4,R)\ltimes R^4$, generated by
$a$, $a^\mu{}_\rho$, and $b_{\mu\nu\rho}$.

%%%%%%%%%%%%%%%%%%%%%%%%%%%%
\subsection{Introduction of Extra Coordinates}
%%%%%%%%%%%%%%%%%%%%%%%%%%%%

In seeking a manifestly realised $SL(5,R)$ symmetry, six extra coordinates 
were introduced in \cite{Duff:1990hn},
such that together with the four coordinates $X^\mu$ of spacetime they 
form a 10-plet of $SL(5,R)$. 
The extra coordinates are antisymmetric tensorial,  and are 
denoted by $Y^{\mu\nu}$. The field equations 
and Bianchi identities of the membrane were cast into a manifestly 
$SL(5,R)$-covariant form. However, the equations 
satisfied by the extended system are problematic, and this can be seen 
as follows.  Setting $B_{\mu\nu\rho}=0$ 
and taking $g_{\mu\nu}=\eta_{\mu\nu}$ for simplicity, these equations take 
the form \cite{Duff:1990hn}
\bea
\begin{split}
\sqrt{-\gamma} \gamma^{ij} \partial_j X^\mu &=& 2 \varepsilon^{ijk} \partial_j Y^{\mu\nu} \partial_k X_\nu\ ,
\\
\sqrt{-\gamma} \gamma^{ij} \partial_j Y^{\mu\nu} &=&  \varepsilon^{ijk} \partial_j X^\mu \partial_k X^\nu\  .
\end{split}
\label{DJXL}
\eea
Taking the curl of the second equation gives the integrability condition
\be
V_i^{\mu\nu}= 0\ ,
\ee
where we have used the field equation $\nabla_i \partial^i X^\mu =0$ to
simplify the result.  This integrability condition, which implies 
second-order differential constraints over and above the field equations,
therefore poses  a problem with the desired $SL(5,R)$ duality-symmetric system of 
membrane equations.  
Note also  in the GZ approach as well as the approach 
in which extra coordinates are introduced, the obstacle to the sought-after $SL(5,R)$ symmetry is the
nonvanishing of the expression $V_i^{\mu\nu}$ defined in \eq{anomaly}. 

A different proposal has been made in \cite{Cederwall:2007je}, where 
the original coordinates are embedded into an $SL(5,R)$ 10-plet $Z^{MN}$ via
\bea
Z^{\mu 5}\equiv X^{\mu}\;,\qquad
Z^{\mu\nu} \equiv Y^{\mu\nu}\;.
\eea
The following ${\rm SL}(5)$ covariant equation was proposed in 
\cite{Cederwall:2007je}:
\bea
\sqrt{-\Gamma} \Gamma^{ij} \partial_j Z^{MN} &=& c\,\varepsilon^{ijk} \,
\partial_j Z^{MP} \partial_k  Z^{NQ} G_{PQ}\;.
\label{master}
\eea
Here $c$ is an arbitrary constant, and the induced ${\rm SL}(5)$-invariant 
metric is given by 
\bea
\Gamma_{ij} &=&  -\frac12\,\partial_i Z^{MN} \partial_j Z_{MN}\ ,
\label{Gamma}
\eea
where $SL(5,R)$ indices are raised and lowered with the metric $G_{MN}$. 
Note the scale invariance of the equation (\ref{master}) under the
rescaling $Z\longrightarrow \lambda Z$\,.
The duality equation (\ref{master}) induces the equations of motion
\bea
\nabla^i \partial_i Z^{MN} &=& 0\;,
\eea
which resemble the original membrane equations of motion, except that
the worldvolume metric $\gamma_{ij}$ is now
replaced by the ${\rm SL}(5)$-invariant metric $\Gamma_{ij}$\,.
The curl of (\ref{master}) also implies the integrability equation
\bea
\nabla_i \partial^j Z^{P[M} \partial_j Z^{N]Q}\,G_{PQ} &=& 0
\;.
\eea
Unlike the previous proposal discussed above, this does not yield 
an immediate contradiction, 
since it involves the original $X^\mu$ as well as the new 
$Y^{\mu\nu}$ coordinates. 
Indeed, taking  $B_{\mu\nu\rho}=0$ and $g_{\mu\nu}=\eta_{\mu\nu}$, 
we see that the equation \eq{master} gives
\bea
\sqrt{-\Gamma} \Gamma^{ij} \partial_j X^\mu &=& 
- c\, \varepsilon^{ijk} \partial_j Y^{\mu\nu} \partial_k X_\nu\ ,
\\
\sqrt{-\Gamma} \Gamma^{ij} \partial_j Y^{\mu\nu} &=&  
c\, \varepsilon^{ijk} \partial_j X^\mu \partial_k X^\nu
+ c\,\varepsilon^{ijk} \partial_j Y^{\mu\sigma} \partial_k Y^\nu{}_\sigma\ .
\eea
Apart from the fact that the induced worldvolume metrics are different, 
we see that the 
second equation above contains an extra term, in comparison to that 
given in \eq{DJXL}. Consequently,
its integrability condition will indeed mix $X^\mu$ and $Y^{\mu\nu}$, 
thereby avoiding an immediate conflict. However, to test whether 
the system 
described by \eq{master} makes sense, we should also examine the spectrum 
of small 
fluctuations around 
a vacuum solution that respects the worldvolume Poincar\'e symmetry.

Such a background can be taken to be given by
\bea
G_{MN}~=~\eta_{MN}\;,\qquad
\partial_i \overline{Z}{}^{MN} &=&
\lambda\,\sigma_i{}^{MN}
\;,
\label{background}
\eea
where $\lambda$ is an arbitrary constant, $\eta_{MN}$ is the
$SO(p,q) \subset SL(5,R)$ invariant tensor,
and $\sigma_i{}^{MN}$ specifies the embedding
of $SO(2,1)$ into $SO(p,q)\subset SL(5,R)$, for which we choose the
canonical normalization
\bea
{}
[\sigma_i, \sigma_j] &=& \varepsilon_{ijk}\,\sigma^k
\;.
\eea
Owing to the scale invariance of (\ref{master}), the 
parameter $\lambda$ drops out, and for convenience we choose it so that we 
may identify the worldvolume metric (\ref{Gamma}) with 
$\eta_{ij}={\rm diag}(-,+,+)$:
\bea
\overline{\Gamma}{}_{ij}
&=&\frac12\,\lambda^2\,{\rm Tr}\,(\sigma_i\sigma_j)~=~\eta_{ij}
\;.
\label{GammaEta}
\eea
Equation (\ref{master}) then yields
\bea
{}[\sigma_i,\sigma_j] &=& \frac1{\lambda c}\,\varepsilon_{ijk}\,\sigma^k\qquad
\Longrightarrow\qquad c\lambda=1
\;.
\eea
Denoting the spin of the representation $\sigma_i$ by $j$
(assuming the representation carries a single spin), we have the relation
\bea
{\rm Tr}\,(\sigma_i\sigma_j) &=& \frac13\,j(j+1)(2j+1)\,\eta_{ij}
\;,
\eea
which together with (\ref{GammaEta}) determines
\bea
c^2 &=& \frac16\,j(j+1)(2j+1)
\label{c2}
\;.
\eea
Defining the fluctuations around this background as
\bea
Z^{MN} &=& \overline{Z}{}^{MN}+ \phi^{MN}
\;,
\eea
we fix the gauge freedom of worldvolume diffeomorphisms by imposing
\bea
\sigma_i{}^{MN}\,\phi_{MN} &=& 0 
\;. 
\label{gauge}
\eea
The expansion of equation (\ref{master}) to linear order in the fluctuations then gives
\bea
\eta^{ij} \partial_j \phi_{MN} +
\varepsilon^{ijk}\,\left(\sigma_{j\,M}{}^P \partial_k \phi_{PN}
-\sigma_{j\,N}{}^P \partial_k \phi_{PM}\right)
&=& 0
\;.
\eea
The final analysis depends on the particular choice of generators $\sigma_i$ embedding 
$SO(2,1)$ into $SL(5,R)$. Three inequivalent choices correspond to the decompositions\footnote{
We do not consider the decompositions $5\rightarrow 2+3$ and $5\rightarrow1+4$ 
because they do not allow for a symmetric invariant $\eta_{MN}$.}
\bea
A) &:&\qquad 5 \rightarrow 5 \;,\nonumber\\
B) &:& \qquad 5 \rightarrow 3+1+1 \;,\nonumber\\
C) &:&  \qquad 5 \rightarrow 2+2+1 \;.
\eea

For case A), using an explicit spin-2 representation for the 
generators $\sigma_i$ implies that
the invariant tensor $\eta_{MN}$ is of signature $(2,3)$, and
one may verify that equation (\ref{master}) reduces to
\bea
\partial_{(i} \phi_{jkl)} &=&0
\;,
\eea
for the components 
$\phi_{ijk}\equiv (\sigma_{(i}\sigma_{j}\sigma_{k)})^{MN}\phi_{MN}$ 
surviving the gauge condition~(\ref{gauge}). 
This shows that around this background the fluctuations do not admit any 
non-trivial dynamics.

Similarly, in case B) equations (\ref{master}) restrict the 
fluctuations to
\bea
\partial_{(i} \phi_{j)4} &=&0\;,\qquad
\partial_{(i} \phi_{j)5} ~=~0\;,\qquad \partial_i \phi_{45}~=~0
\;,
\eea
which again kills all dynamics for the fluctuation components 
surviving the gauge condition~(\ref{gauge}).

Finally, in case C) the background is most conveniently given in terms of 
the 't Hooft symbols
\bea
(\sigma_{i})_M{}^N &=& -\frac{1}{2} \delta_M^{m} \eta^{nN} \varepsilon_{imn} 
+ \frac12\,\eta_{iM} \delta^{N}_4-\frac12\,\delta_{i}^N \eta_{M4} \;.
\eea
(with $\eta_{44}=-1$). In this case, the degeneracy of the 
spin $1/2$ representations introduces
an additional factor of 2 into (\ref{c2}), such that 
$c=\frac1{\sqrt{2}}$\,.
The fluctuation equations from (\ref{master}), together with the gauge 
condition~(\ref{gauge}), imply that
\bea
\partial_i\phi_{45} &=& -\frac12\,\epsilon_i{}^{jk}\,\partial_j \phi_k\;,\qquad
\partial_{(i} \phi_{j)} ~=~ 0\;,
\eea
which again kills all dynamics for the fluctuation components surviving 
the gauge condition.

%%%%%%%%%%%%%%%%
\section{Conclusions}
%%%%%%%%%%%%%%%%

Using the $p$-brane worldvolume approach to U-dualities, we have confirmed 
that there is a problem when $D\neq p+1$, focusing on $p=2$ in $D=4$ 
where the expected $SL(5,R)$ fails to materialise, and $p=2$ in $D=3$, which we 
refer to as the topological membrane, where 
the expected $SL(3) \times SL(2)$ does arise. 
In the case of the topological membrane, we have introduced extra 
coordinates to make the U-duality symmetry manifest.
The features we have found for $D=4$ are the  same whether we use the approach 
where extra coordinates are introduced in 
order to make U-dualities manifest \cite{Duff:1990hn,Cederwall:2007je}, or in the 
Gaillard-Zumino approach where the symmetries are not manifest.
In the latter approach, as well as that of \cite{Duff:1990hn}, we have shown that
the $SL(5,R)$ U-duality fails due to the nonintegrability of the transformation rule of
the pull-back map. In the approach of \cite{Cederwall:2007je}  where a manifestly $SL(5,R)$-invariant 
equation  is proposed for a membrane  in a target based  on generalized geometry, we have shown 
that these equations do not support linearized  fluctuations about a Poincar\' e invariant vacuum solution.

These problems extend
to the worldvolume treatment of  Berman and Perry \cite{Berman:2010is}, 
and also to the approach in   Hatsuda et al.\cite{Hatsuda:2012vm}, which reformulates 
the diffeomorphism  constraints for an M2-brane coupled to a supergravity background in $D=4$  
in an $SL(5,R)$-covariant form.  In both cases the problem is that
they use transformation rules that are not integrable, for the reasons
we have explained above
More specifically, in \cite{Hatsuda:2012vm}, the $SL(5,R)$ transformations
of the time components $J^{0}$ and $P^{0}$ are used to assert the $SL(5,R)$ invariance of the
Hamiltonian constraint, which is a quadratic form in these variables.
By worldvolume Lorentz symmetry, however, also the space components must
transform in the same way, which is equivalent to our \eq{dPJ}.
Then, our discussion leading to the nonintegrability of the resulting transformation
rule for $\partial_i X^\mu$ continues to be an obstacle for $SL(5,R)$ invariance. 
We expect this will also appear in  the case of $SO(5,5)$ symmetry of M5 branes that 
has been proposed in \cite{Hatsuda:2013dya}.

Going beyond $D=5$ only exacerbates the problem, since the U-duality 
multiplets involve the M5-brane charges as well the M2-brane charges  
and the momentum, as shown in the Appendix. 
One possible approach to this problem may be along the lines studied in \cite{Bengtsson:2004nj} for the 
case of M2-brane in $D=8$. In this approach, firstly one keeps all the target space fields arising in the 
dimensional  reduction of  $D=11$ supergravity down to $d$ dimensions. Next, one ensures the gauge 
invariance of the pull-backs of all the  target space form fields by introducing appropriate worldvolume 
potentials, in the same way the worldvolume vector fields are introduced in  D-brane actions. Then, one 
imposes suitable  duality equations that exhibit the expected U-duality symmetry group manifestly, while 
maintaining the correct number of degrees of freedom.  The challenge in this approach is to resolve the 
resulting highly nonlinear constraint equations in a way that maintains the U-duality symmetry. 
So far, these equations have been solved for a restricted class of 
supergravity background such  that only the  two parameter Heisenberg subgroup of $SL(2,R)$ we 
encountered in our treatment of membrane in $D=8$ has been realised \cite{Bengtsson:2004nj}.

The $E_{n(n)}$ symmetry of $D=11$ supergravity compactified on the torus 
$T^n$ also arises if ten-dimensional  type IIA or IIB supergravity is compactified on the torus $T^{n-1}$.  
An alternative approach to finding a  worldvolume realisation of these U-duality symmetries could be to 
look at string theories in $(n-1)$ dimensions  rather than the membrane in $n$ dimensions.  Ideas along 
these lines have been pursued in  \cite{Linch:2015fca,Linch:2015qva,Linch:2015fya,Linch:2015lwa}.

%%%%%%%%%%%%%%%%%%%%%%%%%%%%%
\section*{Acknowledgements} 
%%%%%%%%%%%%%%%%%%%%%%%%%%%%%

We are grateful to Alexandros Anastasiou, Eric Bergshoeff, Leron Borsten, Guillaume Bossard,
Martin Cederwall, Mia Hughes, William Linch, Silvia Nagy, Yi Pang, Daniel Robbins, Andy Royston, 
Arkady Tseytlin and Peter West for discussions. JXL acknowledges support by a key grant from 
the NSF of China with  Grant No : 11235010.  HS thanks Mitchell Institute for Fundamental Physics 
and Astronomy at  Texas A\& M University, and ES thanks Lyon 
University, and CNP and ES thank the Mitchell Foundation for 
hospitality at the Great Brampton House Workshop on
Cosmology and Strings, April 2015, where part of this work was carried out. 
CNP is supported in part by DOE grant DE-FG02-13ER42020, and 
ES is supported in part  by NSF grant PHY-1214344.

%%%%%%%%%%%%%%%%%%%%%%%
\begin{appendix}
%%%%%%%%%%%%%%%%%%%%%%%
\section{Supergravity Compactifications}
\label{sec:compact}
%%%%%%%%%%%%%%%%%%%%%%%

%\subsection{$\mathcal{N}=1, D=11$}

Recall that the 11=dimensional M-theory superalgebra is given by \cite{Townsend:1995gp} :
\begin{equation}
\label{eq:MAlgebra}
\{Q_\alpha,Q_\beta\} = \big(\Gamma^MC\big)_{\alpha\beta}P_M 
 \big(\Gamma^{MN}C\big)_{\alpha\beta}\, Z_{MN} + \big(\Gamma^{MNPQR}C\big)_{\alpha\beta}\, Z_{MNPQR}.
\end{equation}
where $Q_\alpha$ transforms as $\rep{32}$ of $SO(10,1)$. The total number of components of all 
charges on the RHS is
\begin{equation}
\label{eq:charges}
\rep{11} + \rep{55} + \rep{462} = \rep{528},
\end{equation}
which is, algebraically, the maximum possible number since the LHS is a symmetric $32 \times 32$ matrix. 
The spatial components of the momentum  $  P_1$ and the central charges $Z_2, Z_5$ are associated with 
the plane wave W1, the 2-brane M2 and the 5-brane M5;  the temporal components are associated with their 
duals, the KK-monopole K6, and objects we can call K9 and W10.

After dimensional reduction to n dimensions, the 528 charges will form representations of 
$SO(n-1,1)\times SO(D)$ as in (\ref{compact}).

\be
%\begin{table}[H]
\begin{array}{cccccccccc}
%\caption{$p$-form charges in $SO(D-1,1)$ representations}
%\begin{ruledtabular}
%\begin{tabular*}{\textwidth}{*{16}{M{c}}}
&&p                     &0   & 1                & 2                    & 3                   & 4               & 5               &\\
&n=11-D&&&&&&&&\\
%\hline
& 11  &&          & \rep{11}       & \rep{55}      &    &      & \rep{462}  &\\
& 10  & &\rep{1}           & \rep{10 + 10}       & \rep{45}      &  &  \rep{210}      & \rep{252}  &\\
& 9  && \rep{(1,3)}           & \rep{(9,3)}        & \rep{(36,1)}         & \rep{(84,1)}        & \rep{(126,3)}       & &\\
& 8   && \rep{(1,6)}           & \rep{(8,4)}        & \rep{(28,2)}         & \rep{(56,4)}        & \rep{(70,3)}     &&\\\
& 7   && \rep{(1,10)}          & \rep{(7,6)}        & \rep{(21,6)}         & \rep{(35,10)}            & \\
& 6   && \rep{(1,16)}          & \rep{(6,12)}       & \rep{(15,16)}        & \rep{(20,10)}              & &  \\
& 5   &&  \rep{(1,28)}         & \rep{(5,28)}       & \rep{(10,36)}        &                 && \\
& 4   &&  \rep{(1,56)}         & \rep{(4,64)}       & \rep{(6,36)}         && \\
& 3   &&  \rep{(1,120)}        & \rep{(3,136)}      &               &\\
& 2   &&  \rep{(1,256)}        & \rep{(2,136)}      &                    &\\
& 1   &&   \rep{528}      &       &                       &\\
%\end{tabular*}
%\end{ruledtabular}
\label{compact}
%\end{table}
\end{array}
\ee

However, the charges carried by the waves, branes and monopoles do not fall into representations of $SO(n-1,1)\times SO(D)$ because they discriminate between the temporal and spatial components. For example, writing $M=(0,I)$ with $I=1,2,...10$, the 45 $SO(10)$ M5 charges are given by  $Z_{IJ}$ and the 10 K9 charges by $Z_{0J}$ or equivalently ${\tilde Z }_{IJKLMNOPQ}$.  These $SO(n-1)\times SO(D)$ reps are given in ( \ref{pforms}).
\be
\begin{array}{cccccccccccccccc}
%\begin{table}[H]
%\caption{Wave, brane and monopole charges and how they transform under $SO(D-1)$}\label{tab:pforms}
%\begin{ruledtabular}
%\begin{tabular*}{\textwidth}{*{16}{M{c}}}
p                      &0   & 1                & 2                    & 3                   & 4               & 5               &      6   & 7&8&9&10\\
n=11- D&&&&&&&\\
%\hline
 11  &          & \rep{10}       & \rep{45}         &      &  & \rep{252}   & \rep{210}   &   &   & \rep{10}      & \rep{1}&\\
 10  & \rep{1}           & \rep{9+9}        & \rep{36}         &      & \rep{126}   & \rep{126+126}   & \rep{84}    &   & \rep{9}     & \rep{1+1}       & &\\
 9   & \rep{(1,3)}           & \rep{(8,3)}        & \rep{(28,1)}         & \rep{(56,1)}        & \rep{(70,3)}    & \rep{(56,3)}    & \rep{(28,1)}   &  \rep{(8,1)}     & \rep{(1,3)}   &&\\\
 8   & \rep{(1,6)}           & \rep{(7,4)}        & \rep{(21,2)}         & \rep{(35,4)}        & \rep{(35,6)}    & \rep{(21,4)}    & \rep{(7,2)}     & \rep{(1,4)}     & \\
7   & \rep{(1,10)}          & \rep{(6,6)}        & \rep{(15,6)}         & \rep{(20,10)}       & \rep{(15,10)}   & \rep{(6,6)}     & \rep{(1,6)}        & &  \\
 6   &  \rep{(1,16)}         & \rep{(5,12)}       & \rep{(10,16)}        & \rep{(10,20)}       & \rep{(5,16)}    & \rep{(1,12)}    &                 && \\
 5   &  \rep{(1,28)}         & \rep{(4,28)}       & \rep{(6,36)}         & \rep{(4,36)}        & \rep{(1,28)}    &                 &                 && \\
 4   &  \rep{(1,56)}         & \rep{(3,64)}       & \rep{(3,72)}         & \rep{(1,64)}        &                 &                 &                  &\\
 3   &  \rep{(1,120)}        & \rep{(2,136)}      & \rep{(1,136)}        &                     &                 &                 &                 &\\
2   &  \rep{(1,256)}        & \rep{(1,272)}      &                      &                     &                 &                 &                 &\\
 1   &  \rep{(1,528)}        &                    &                      &                     &                 &                 &\\
%\end{tabular*}
%\end{ruledtabular}
%\end{table}
\label{pforms}
\end{array}
\ee
Note that only the 0-brane charges can be assigned to a representation (fundamental) of the non-compact U-duality 
as opposed to its maximal compact subgroup. In $D=3$, for example, we have the $(3,2)$ of $SL(3)\times SL(2)$ 
with generalised coordinates
\be
Z^M=( X^{\mu},Y_{\rho\sigma})~~~\mu=1,2,3
\ee
In $D=4$ we have the $10$ of $SL(5)$ with generalised coordinates
\be
Z^M=( X^{\mu},Y_{\rho\sigma})~~~\mu=1,2,3,4
\ee
In $D=7$ we have the $56$ of $E_{7(7)}$ with generalised coordinates
\be
Z^M=( X^{\mu},Y_{\rho\sigma},{\tilde Y}_{\lambda\tau},{\tilde X}^{\nu}) ~~~\mu=1,2,3,4,5,6,7
\ee

\end{appendix}

\providecommand{\href}[2]{#2}\begingroup\raggedright\endgroup

\end{document}